\documentclass[letterpaper,11pt]{article}
\pdfoutput=1 
\usepackage{amsmath}
\usepackage{amsthm}
\usepackage{amsfonts}

\usepackage{algorithm}
\usepackage{algpseudocode}
\usepackage{comment}
\usepackage{amsthm}
\usepackage{algpseudocode}
\usepackage{amssymb}
\usepackage{algorithm}
\usepackage{graphicx}
\usepackage{epstopdf}
\DeclareGraphicsExtensions{.eps}
\usepackage[export]{adjustbox}
\usepackage{color}
\usepackage{float}
\usepackage{pgfplots}
\usepackage[normalem]{ulem}
\usepackage{fullpage}
\usepackage{algpseudocode}
\usepackage{authblk}
\usepackage{subfigure}
\usepackage{multirow}

\date{}
\makeatletter
\newcommand\footnoteref[1]{\protected@xdef\@thefnmark{\ref{#1}}\@footnotemark}
\makeatother
\newlength\figureheight \newlength\figurewidth
\newcommand{\mytilde}{\raise.17ex\hbox{$\scriptstyle\mathtt{\sim}$}}
\title{ComHapDet: A Spatial Community Detection Algorithm for Haplotype Assembly}

\author{%
	Abishek Sankararaman \footnote{\noindent Electrical and Computer Engineering, The University of Texas at Austin. Email - abishek@utexas.edu},\hspace{4mm}
	Haris Vikalo \footnote{Electrical and Computer Engineering, The University of Texas at Austin. Email - hvikalo@ece.utexas.edu},\hspace{4mm}
	Fran\c cois Baccelli, \footnote{Dept of Mathematics and Electrical and Computer Engineering, The University of Texas at Austin.  Email - baccelli@math.utexas.edu}
}

\begin{document}
	\maketitle

%\begin{abstractbox}
\begin{abstract}
	\noindent
{\textbf{Background:} Haplotypes, the ordered lists of single nucleotide variations that distinguish chromosomal sequences 
from their homologous pairs, may reveal an individual's susceptibility to hereditary and complex diseases 
and affect how our bodies respond to therapeutic drugs. Reconstructing haplotypes of an individual from 
short sequencing reads is an NP-hard problem that becomes even more challenging in the case of polyploids. 
While increasing lengths of sequencing reads and insert sizes {\color{black} helps improve accuracy of reconstruction}, 
it also exacerbates computational complexity of the 
haplotype assembly task. This has motivated the pursuit of algorithmic frameworks capable of accurate
yet efficient assembly of haplotypes from high-throughput sequencing data. \\
\noindent \textbf{Results:} We propose a novel graphical representation of sequencing reads and pose the haplotype 
assembly problem as an instance of community detection on a spatial random graph. To this end, we 
construct a graph where each read is a node with an unknown community label associating the read with
the haplotype it samples. Haplotype reconstruction can then be thought of as a two-step procedure: first, 
one recovers the community labels on the nodes (i.e., the reads), and then uses the estimated labels to 
assemble the haplotypes.  Based on this observation, we propose {\ttfamily ComHapDet} -- a novel assembly 
algorithm for diploid and ployploid haplotypes which allows both bialleleic and multi-allelic variants. \\
\noindent \textbf{Conclusions:}  Performance of the proposed algorithm is benchmarked on simulated as well as experimental data 
obtained by sequencing Chromosome $5$ of  tetraploid biallelic \emph{Solanum-Tuberosum} (Potato).
The results demonstrate the efficacy of the proposed method and that it compares favorably with the existing
techniques. 

% \url{https://github.com/abishek90/ComHapDet-Repo}

%	\textbf{Contact:} \href{name@bio.com}{name@bio.com}\\
%	\textbf{Supplementary information:} Supplementary data are available at \textit{Bioinformatics} online.
}

\end{abstract}
%
%\begin{keyword}
%	\kwd{Haplotype Assembly}
%	\kwd{Spatial Random Graph}
%	\kwd{Graph Clustering}
%	
%\end{keyword}
%
%\end{abstractbox}
%%
%%\end{fmbox}% uncomment this for twcolumn layout
%
%\end{frontmatter}
%%\maketitle

\section{Introduction and Background }

Technological advancements in DNA sequencing have enabled unprecedented studies of genetic blueprints and variations between
individual genomes. An individual genome of a eukaryotic organism is organized in $K$-tuples of homologous chromosomes; diploids
($K=2$), including humans, have genomes organized in pairs of homologous chromosomes where the chromosomes in a pair differ 
from each other at a small fraction of positions. The ordered lists of such variants -- the so-called single nucleotide polymorphisms 
(SNPs) -- are referred to as \emph{haplotypes}. Many plants are polyploid, i.e., have genomes organized in $K$-tuples, $K > 2$, of 
homologous chromosomes; for instance, the most commonly grown potato crop (\emph{Solanum Tubersoum}) is a tetraploid 
organism with a total of $48$ chromosomes (i.e., $12$ distinct quadruplets). Haplotype information of an individual is useful in a wide 
range of applications. For 
instance, in humans, the haplotype information contains indicators to the individual's susceptibility to diseases and expected responses 
to certain classes of drugs \cite{haplo_role_humans}. Haplotype sequences are also used to infer recombination patterns and identify 
genes under positive section \cite{gene_positive_selection}. In the case of agricultural crops such as the tuber family, the haplotypes 
provide insight into developing disease and pest resistant variety of crops \cite{potato_nature}. Thus, fast and accurate algorithms for 
both polyploid and diploid haplotype reconstruction (or also known as phasing) from high throughput sequencing reads are highly desirable.
\\

Recent advancements in DNA sequencing technologies have dramatically improved affordability and speed of sequencing; the most
widely used are high-throughput sequencing devices (e.g., the Illumina's platforms). Information provided by such platforms
typically comes in form of paired-end reads, each consisting of two short contiguous samples of the chromosome
(separated by a
few hundred bases). Typical reference-guided data processing pipeline starts by determining the relative ordering of the reads via mapping 
them to a reference genome; this step is followed by the detection of variant positions and \emph{SNP calling}. To perform
haplotype assembly, existing methods typically retain only the reads that cover variant positions; parts of the retained 
reads that cover homozygous sites are also discarded. Then the remaining information needs to be used to identify $K$ distinct 
haplotype sequences $S_1,\cdots,S_K$, {\color{black}of same length}. From the previous discussion, it is clear that there is no position in these $K$ strings where all
the alleles are identical (such a position would be homozygous and therefore discarded in the pre-processing step). While the relative
positions of reads are revealed by mapping them onto a reference, their origin remains unknown, i.e., it is not known which among the $K$
chromosomes a given read samples. In the absence of sequencing errors, grouping reads according to their origin is rather
straightforward and based on the disagreement regarding the allele information that the reads provide for each 
variant site. Unfortunately, however, sequencing is erroneous with state-of-the-art sequencing platforms achieving sequencing errors 
in the range of $10^{-3} - 10^{-2}$. In the presence of errors, it is no longer obvious how to decide whether a read originates from 
a particular haplotype; to this end, one needs to rely on a computational framework for haplotype assembly -- the central focus of the 
present paper.
\\

Existing work formulates haplotype assembly as a combinatorial optimization problem where one seeks to reconstruct the haplotype 
sequence by judiciously making as few modifications of the data as possible in order to remove read membership ambiguities
arising due to sequencing errors. This has led to optimization criteria such as minimal fragment removal, minimum SNP removal 
\cite{MEC_Hard1}, maximum fragments cut \cite{refhap} and minimum error correction (MEC) \cite{mec_algo}. Motivated by the
observation that the MEC score optimization is NP-hard \cite{MEC_Hard1,MEC_Hard2}, a considerable amount of recent work 
studied relaxations of the underlying combinatorial optimization problem \cite{h_pop,hapcol,probabilistic_haplotyping,MEC_Hard2}.
In a pioneering work, \cite{greedy} proposed a greedy algorithm aimed at assembling most-likely haplotype sequences given the 
observations. This line of thought of using Bayesian methods to reconstruct the most likely haplotype sequence was carried out 
further in \cite{mcmc_1} and \cite{mcmc_2} using MCMC and Gibbs sampling, respectively. However, these methods are usually 
slow as the associated Markov chains have large mixing times, thereby making their practical applicability limited. HapCUT 
\cite{hapcut} makes a connection between haplotype assembly and graph-clustering, and solves it by identifying a maximum cut in 
an appropriately constructed graph. This method was shown to be superior to \cite{greedy} and is widely used in benchmarking. 
HapCUT algorithm was then significantly outperformed by HapCompass \cite{hap_compass}. However, apart from HapCompass,
all of these methods are restricted to the diploid case. To address both the diploid and polyploid scenarios, SDhaP \cite{sdhap} 
relaxes the MEC minimization problem to a convex optimization program and solves it efficiently by exploiting the underlying data
structure. More recently, AltHap \cite{vikalo_tensor} casts haplotype assembly as a low-rank tensor factorization problem and
solves it via fast alternating optimization heuristics. The connection to matrix factorization was previously exploited in
\cite{matrix_factorization_haplo} and \cite{matrix_2}, but those methods were incapable of handling polyploids or polyallelic 
assembly problem. Prior to the current paper, AltHap \cite{vikalo_tensor} is the only algorithm capable of solving the
assembly problem in the polyploids/polyallelic scenario.

\subsection*{Main Contributions}

In this paper, we propose a novel formulation of haplotype assembly as a \emph{spatial graph} clustering problem. This formulation
is based on \emph{spatial point process} representation of paired-end reads; in particular, we argue that assigning to each 
paired-end read spatial coordinates corresponding to the starting indices of the reads in the pair represents valuable augmentation 
of the information. Equipped with such a representation of the data, we construct a graph whose nodes represent the reads, and
place a weighted edge between two reads if they overlap in at least one position. The edge weights are formed using the scoring 
function adopted from \cite{sdhap}; this weighting mechanism ensures that if two reads belong to the same haplotype, then the
edge connecting them will likely be assigned a large positive weight, while if they belong to different haplotypes, then the edge 
between them will likely be assigned a large negative weight. We then cast the haplotype assembly problem as an Euclidean 
community detection problem \cite{com_det}, where the community label of a node (i.e., a read) indicates the haplotype  
 it comes from. We find in our experiments that such a `spatial' embedding of the problem greatly improves both the accuracy 
and the run time complexity of polyploid phasing. The improvement in accuracy stems from the fact that our algorithm naturally 
enforces a \emph{regularizing constraint} that the underlying sampling process is in a sense uniformly distributed in space. In 
other words, we know that in every `location' of space, the total number reads covering the given SNP location belonging to the various haplotypes are identical.
Having a spatial representation of the data allows one to incorporate this prior knowledge about the 
sampling process and leads to dramatically higher accuracy, especially in the polyploid case. The 
spatial representation is also crucial in reducing run times, as this exposes the problem's inherent `locality' condition in space. 
More precisely, observe that if two reads are `far away' in this embedding (i.e, they do not overlap), there will be no edge 
connecting them in the corresponding graph. This allows us to naturally `decompose' the haplotype assembly problem using 
a divide-and-conquer paradigm, where we can perform assembly on smaller spatially localized sets of reads -- which is 
computationally very efficient -- and then perform a synchronization step to combine the local haplotype assemblies into a single 
global solution. This algorithmic framework is robust to inaccuracies in the individual local instances of phasing since the 
synchronization or the combining step has a natural \emph{error-correction} mechanism due to a single location being phased in 
multiple local instances. Thus, we can employ faster but less accurate local phasing methods while still achieving high global 
accuracy and good run-time performance. Further technical aspects of these ideas are elaborated in Section \ref{sec:algorithm}.

\section{Problem Formulation}
\label{sec:model}

\subsection*{Setup and Notation}

{\color{black}
Let $m$ and $n$ denote the length of the haplotype sequences and the total number of paired-end read measurements, respectively.
Let $k$ denote the ploidy, i.e., the number of haplotype sequences and $a$ be 
the cardinality of the alphabet set. If the variant sites are polyallelic, then $a=4$ (i.e., all $4$ alleles $A,C,G,T$ are possible); in diploid
and polyploid bi-allelic case, $a=2$. The haplotype sequences are denoted by $s_l[i]$, where for each $i \in \{1,\cdots,m\}$ and 
$l \in \{1,\cdots,k\}$, we have $s_l[i] \in \{1,\cdots,a\}$. As an example, in the poly-allelic case, we have  $s_l[i] \in \{\text{A},\text{C},\text{G},\text{T}\}$. In the rest of 
the paper, we refer to the haplotype positions $\{1,\cdots,m\}$ as \emph{sites}. 
\\

Each read $r_u$, $u \in \{1,\cdots,n\}$, is formed by first
 sampling a haplotype indexed by $v_u \in \{1,\cdots,k\}$ and then sampling a sequence of alleles 
$\{\tilde{s}^{(u)}[i]\}_{i \in \mathcal{I}_u}$ at sites $\mathcal{I}_u \subset \{1,\cdots,m\}$, where for each $i \in \mathcal{I}_u$, 
$\tilde{s}^{(u)}[i] \in \{A,C,G,T\}$ is a ``noisy" (due to sequencing errors) version of the underlying ground truth $s_{v_u}[i]$. 
For each measurement $u \in \{1,\cdots,n\}$, we observe the set of positions $\mathcal{I}_u$ and the noisy values 
$\{\tilde{s}^{(u)}[i]\}_{i\in \mathcal{I}_u}$ but not the index $v_u$ of the haplotype from which the read originates. Thus, using the above notation, the total dataset is denoted as $(\{\tilde{s}^{(u)}[i]\}_{i\in \mathcal{I}_u})_{u \in \{1,\cdots,n\}}$. The goal of 
assembly is to infer origins of the reads and recover haplotype sequences.

\begin{figure}
	\centering
	\includegraphics[scale=0.15]{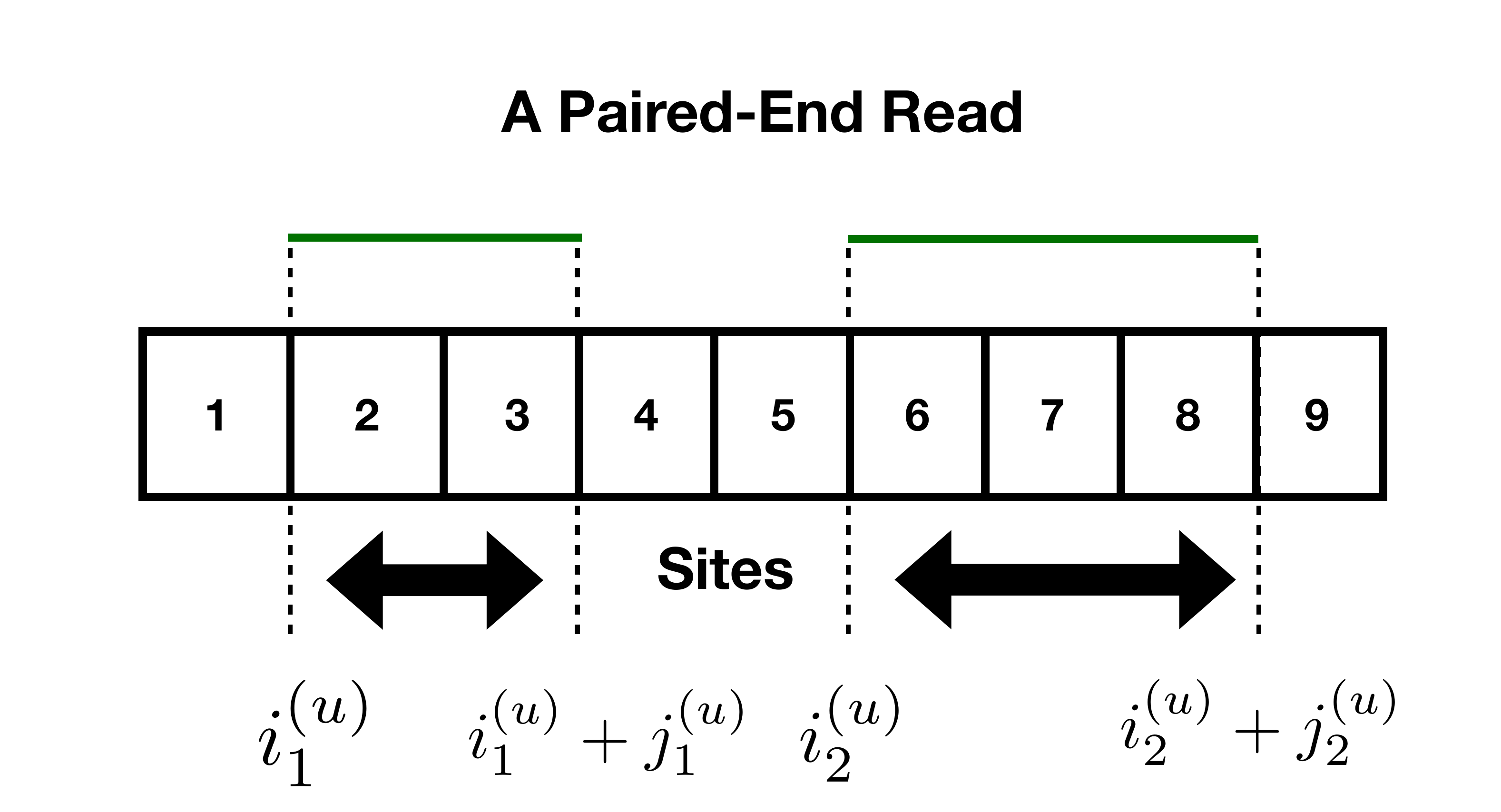}
	\caption{A pictorial description of a paired end read on an example with $m=9$.}
	\label{fig:paired-end}
\end{figure}

In this paper, we consider the case where the data is in form of \emph{paired-ended reads}. Formally, this implies that each measurement 
$u \in \{1,\cdots,n\}$ is such that the set of sites covered by read $u$ has \emph{two contiguous} blocks. More precisely, we assume 
that each read $u \in \{1,\cdots,n\}$ is such that there exists $i_1^{(u)}, j_1^{(u)}, i_2^{(u)},j_2^{(u)} \in \{1,\cdots,m\}$, such that the set of sites covered by $u$, which was denoted by $\mathcal{I}_u$ satisfies,
$\mathcal{I}_u = \{i_1^{(u)},i_1^{(u)}+1,\cdots,i_1^{(u)}+j_1^{(u)}\}  \cup \{i_2^{(u)},i_2^{(u)}+1,\cdots,i_2^{(u)}+j_2^{(u)}\}$. See Figure \ref{fig:paired-end} for an illustration.
\\

\begin{figure}
	\centering
	\includegraphics[scale=0.15]{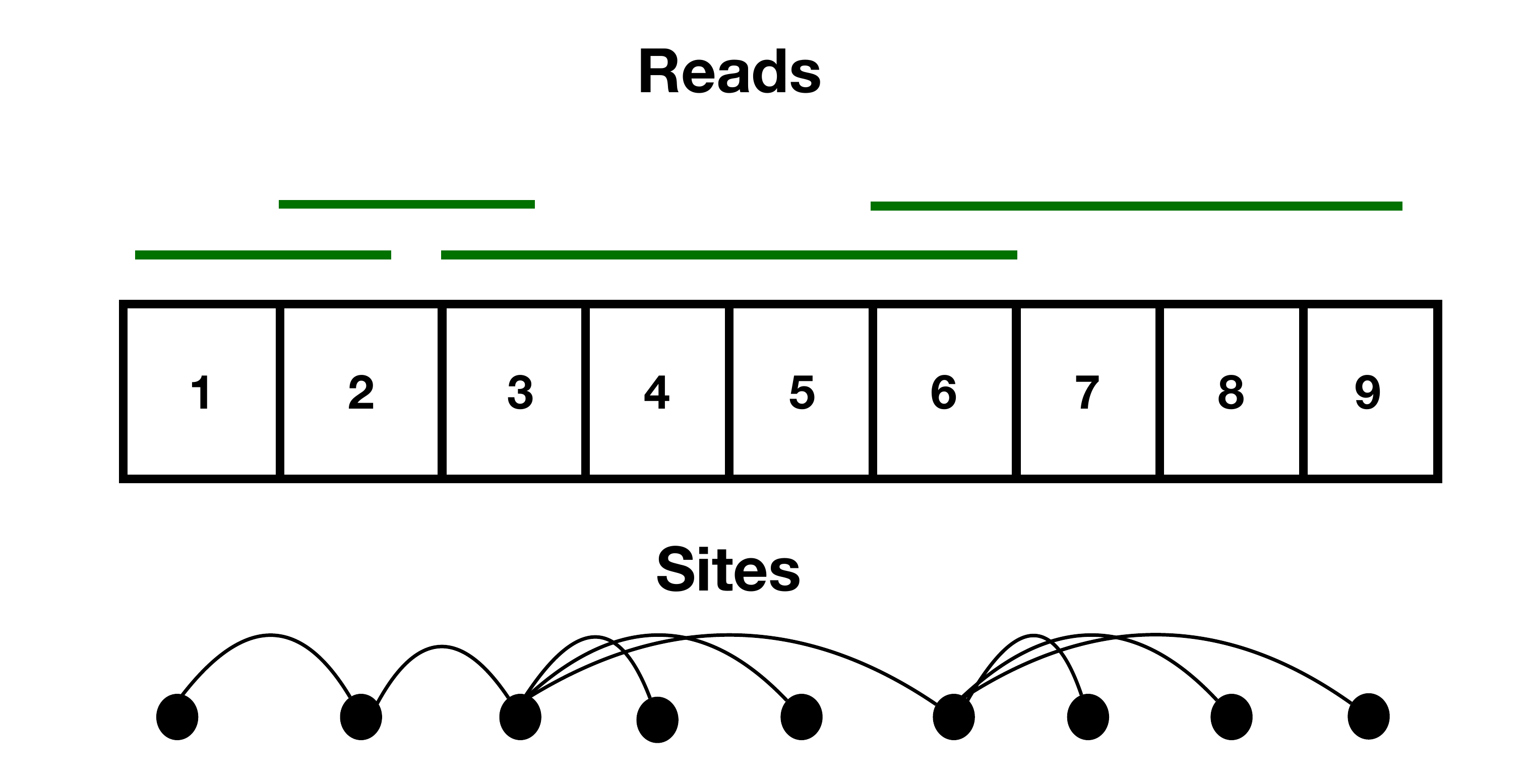}
	\caption{A pictorial description of a connected block of sites. There are $4$ reads in the figure corresponding to the green bars. They set of sites covered by them are $\{1,2\}$, $\{2,3\}$, $\{3,4,5,6\}$ and $\{6,7,8,9\}$ respectively. The bottom figure is the graph $Q$ on $m=9$ vertices constructed by the $4$ reads.}
	\label{fig:contiguous}
\end{figure}

We further assume that the set of $m$ sites and $n$ reads forms a single \emph{connected component}.  In particular, if we draw a graph $Q$ on the vertex set $\{1,\cdots,m\}$, where for any two $a,b \in \{1,\cdots,n\}$, there is an edge in $Q$ if and only if there is a read $u \in \{1,\cdots,n\}$, that covers both sites $a$ and $b$. We say that our data set consisting of the $m$ sites and $n$ reads is \emph{connected}, if the graph $Q$ has a single connected component. Refer to Figure \ref{fig:contiguous} for an illustration.  Notice that since the problem definition is agnostic to how we label the strings 
from $\{1,\cdots,k\}$, any haplotype phasing algorithm can only hope to recover the $k$ strings up to a permutation of the string label (which takes values in $\{1,\cdots,k\}$).
For two disjoint blocks of a haplotype that are not bridged by a read, there is no way to ascertain how to phase fragments of the $k$ 
haplotypes locally recovered inside the blocks. For this reason (and without the loss of generality of the proposed assembly
framework), we assume that the set of haplotypes form a {contiguous block} of reads. If this were not the case, we could pre-process 
the reads and split the problem into many smaller instances, where each instance consists of a single contiguous haplotype block 
that needs to be assembled independently of other blocks.

}

\subsection*{Recovery Goals and Performance Metrics}
 
In this subsection, we formalize the main performance metrics used to benchmark haplotype assembly algorithms, namely the Correct 
Phasing Rate (CPR) and the Minimum Error Correction (MEC) score (see eg. (\cite{exact_whole_genome}, \cite{optimal_whole}, 
\cite{diploid_simulated_benchmark})). The CPR measures the discrepancy between the reconstructed haplotypes 
$\hat{s}_1,\cdots,\hat{s}_k$ and the ground truth $s_1,\cdots,s_k$,
 \begin{align}
 \text{CPR} = \frac{1}{m} \sum_{i=1}^{m} \max_{\pi \in \mathcal{S}_k} \prod_{l=1}^{k} \mathbf{1}_{\hat{s}_l[i] = s_{\pi(l)}[i]},
 \label{eqn:CPR_strong}
 \end{align}
 where $\mathcal{S}_k$ is the set of all permutations of $\{1,\cdots,k\}$. Note that this is a more demanding notion of recovery compared 
 to that used in \cite{vikalo_tensor} and \cite{tripoly}; the metric used there, so-called Reconstruction Rate, we for convenience denote by 
 ${\text{M-CPR}}$ (abbreviating ``Modified CPR"). This metric is defined as
 \begin{align}
{\text{M-CPR}} =  \max_{\pi \in \mathcal{S}_k}\frac{1}{mk} \sum_{i=1}^{m} \sum_{l=1}^{k} \mathbf{1}_{\hat{s}_l[i] = s_{\pi(l)}[i]}.
  \label{eqn:cpr_weak}
 \end{align}
In the case of assembly of diploid haplotypes (i.e., $k=2$), $\text{CPR}$ and ${\text{M-CPR}}$ are identical. However, in the polyploid case 
where the size of the alphabet is generally $4$, it holds that $\text{CPR} \leq {\text{M-CPR}}$ since for all sets $X_1,\cdots,X_k$, 
$\prod_{j=1}^{k} \mathbf{1}_{X_j} \leq\sum_{j=1}^{k} \mathbf{1}_{X_j}$. We adopt $\text{CPR}$ in equation (\ref{eqn:CPR_strong}) since it 
reflects the fraction of sites where the haplotype phasing algorithm correctly recovers all the haplotypes. Unfortunately, the design of algorithms for direct minimization 
of this objective is infeasible since the ground truth is unknown. A commonly used proxy metric in practice is the MEC which can be 
computed directly from the observed data and the reconstructed output. The MEC score is defined as
\begin{align*}
\text{MEC} = \sum_{u=1}^{n} \min_{l \in \{1,\cdots,k\}}\sum_{i=1}^{m} \mathbf{1}_{\text{Read }u \text{ covers site }i} \mathbf{1}_{\hat{s}^{(u)}[i] \neq s_l[j]}.
\end{align*}
 A number of existing haplotype assembly schemes such as \cite{vikalo_tensor}, \cite{hapcut}, attempt to directly minimize the MEC score 
 by solving relaxations of the underlying combinatorial optimization problem that is known to be NP-hard \cite{MEC_Hard1}, \cite{MEC_Hard2}. 
 Contrary to this common approach, in this paper we do not attempt to directly minimize the MEC but rather leverage inherent structural 
 properties of the data and the fact that the noise in measurements is introduced randomly, rather than by an adversarial source, to design 
 a randomized assembly algorithm. Specifically, we rely on the above observations to provide a solution to haplotype assembly by posing it 
 as a clustering problem on a random graph. 
 \\

The key parameters that impact the performance of assembly are \emph{coverage}, \emph{error rate} and \emph{effective read length}. 
Formally, effective read-length $R$ is defined to be the average of $j_1^{(l)}$ and $j_2^{(l)}$, i.e.,
\begin{align*}
R = \frac{1}{2n} \sum_{u=1}^{n} (j_1^{(u)}+j_2^{(u)}).
\end{align*}
We define the coverage $\mathcal{C}$ as the average number of reads that cover a single base in a haplotype, i.e., $\mathcal{C} = \frac{2nR}{km}$. 
Since there are $n$ reads, each covering on average $2R$ haplotype alleles, the total average number of bases being read is $2nr$. The error-rate 
$p$ is the average error rate of the data acquisition process, i.e., the fraction of alleles incorrectly represented by the reads; this rate is aggregate 
of the sequencing and SNP calling errors. We adopt the standard practice of characterizing and benchmarking the performance of haplotype assembly 
algorithms using either the achieved MEC in practical settings where the ground truth is unknown, and the CPR in simulation studies where the 
ground truth is known. We will characterize the performance of our algorithm in settings with varied ploidy, alphabet size, coverage, read-lengths 
and error-rates. 

%\begin{methods}

\section{ The Haplotype Assembly Algorithm}
\label{sec:algorithm}

The algorithm we propose is based on identifying a simple connection between the aforementioned haplotype reconstruction problem and Euclidean 
community detection. Although such a connection was previously noted in the special case of single-ended reads and the diploid haplotype phasing 
problem \cite{tse_comm_det}, no prior work explored this connection in the case of paired-end reads and phasing polyploids. For the first time, we 
provide a unified framework based on Euclidean community detection (e.g., \cite{com_det},\cite{abbe_CD}) for both diploid and polyploid haplotype phasing problems.

 \subsection*{Pre-Processing the Data}

In order to invoke a connection to spatial community detection, we pre-process the $n$ paired-end reads into a graph $G$ with $n$ nodes, where each 
node represents a paired-end read. This pre-processing has two steps - {\em (i)} Constructing weights between pairs of nodes (reads), {(ii)} Placing \emph{labels} on the nodes.
\\

\noindent
{\bf 1. Weights between nodes} - For any two reads $u,v \in [n]$ with $u \neq v$, denote the intersection of sites at which the two measurements 
occur by $\mathcal{I}_u \cap \mathcal{I}_v:=\{l_1,\cdots,l_q\}$, where $q=0$ implies empty set. More precisely, each $l_i$, for $i \in \{1,\cdots,q\}$, is a 
position along the haplotype covered by both $u$ and $v$. If $q=0$, reads $u$ and $v$ cover disjoint set of sites; in this case, there is no edge between 
$u$ and $v$ in the graph $G$. If on the other hand $q > 0$, then we place an edge between them and assign it weight $w_{uv}$ given by
\begin{equation}
w_{uv} := \frac{ 1 }{q} \sum_{h=1}^{q} (\mathbf{1}_{\tilde{s}^{(u)}[l_h] = \tilde{s}^{(v)}[l_h]} - \mathbf{1}_{\tilde{s}^{(u)}[l_h] \neq \tilde{s}^{(v)}[l_h]}  ).
\label{eqn:weights_defn}
\end{equation}

In words, the weight of an edge between any two overlapping reads $u$ and $v$  is the difference between the number of positions (sites) where $u$ and $v$ agree and the number where they disagree, divided by the total number of sites in common. Observe in the definition that
the weights $w_{uv} \in [-1,1]$ for all $u,v \in [n]$. Such a weighting scheme ensures that if $w_{uv}$ is positive and large, then it is \emph{likely} that the reads 
$u$ and $v$ are generated from the same string, while if $w_{uv}$ is negative and large in magnitude, then it is likely that the reads $u$ and $v$ are 
generated from different strings. This \emph{bias} in the weights $w_{uv}$  can be understood by examining the \emph{typical structure} of the polyploid phasing problem. 
Note that if the SNP positions were called accurately, i.e., all of the $m$ haplotypes to be phased were `true', then it would hold that in any location 
$i \in \{1,\cdots,m\}$, not all strings $s_1,\cdots,s_k$  have identical bases, i.e., the set of locations $\{i \in \{1,\cdots, m\} : s_1[i] = \cdots = s_k[i]\} 
= \emptyset$. Since sequencing errors are `typically' small, it is thus the case that if two reads covering the same site have different values, then it is 
\emph{likely} that they come from different haplotypes. 
\\

\begin{figure}
	\centering
	\includegraphics[scale=0.2]{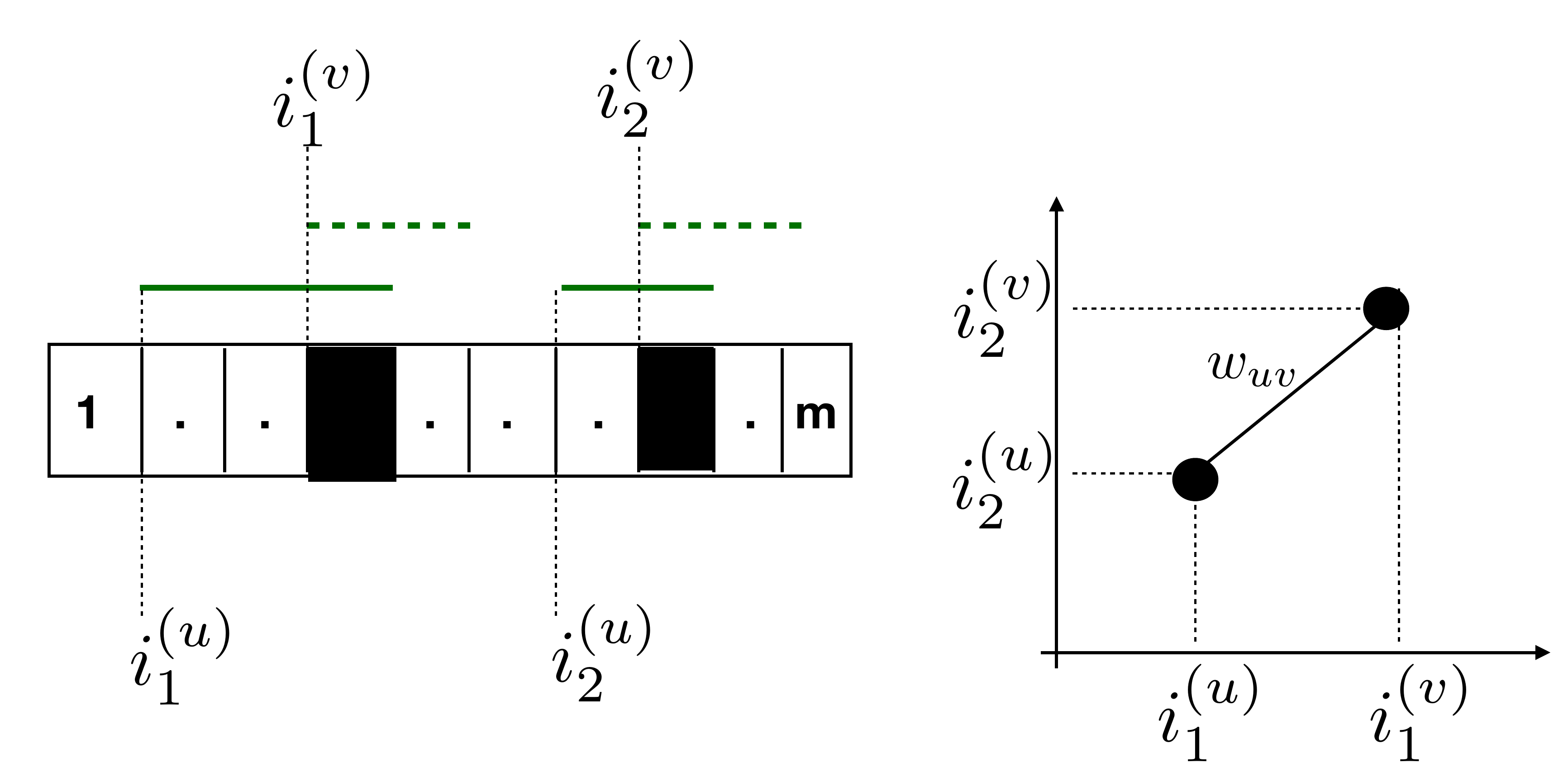}
	\caption{An illustration of our pre-processing the set of reads into a spatial random graph. The starting indices of the two contiguous blocks of a read forms its spatial coordinate and the weight is computed using Equation \ref{eqn:weights_defn}. In this example, the shaded sites contribute to the weight $w_{uv}$.}
	\label{fig:graph_rep}
\end{figure}

\noindent
{\bf 2. Node Labels} - To each node $u \in [N_n]$ of the graph we assign two labels: {\em (i)} - a \emph{community label} $Z_u \in [k]$ and {\em (ii)} - a \emph{spatial label} $X_u \in [m]^2$. The 
community label of a node indicates origin of the corresponding read (i.e., which haplotype the read samples), while the spatial label of a node
$u \in [n]$ is $(i^{(u)}_1,i^{(u)}_2)$, i.e., the locations along the haplotype where the two contiguous substrings of the paired-end read start. See Figure \ref{fig:graph_rep} for an illustration of the data pre-processing. This 
definition of spatial labels assumes that the reads consist of precisely two contiguous substrings; however, experimental data show that this may be 
violated due to various practicalities in base-calling, read mapping and variant calling steps (e.g., due to omitting bases with confidence scores below a 
threshold); consequently, in practice a read (more precisely, the parts 
of a read covering heterogeneous variant positions that are being used in haplotype assembly) may consist of either a single or more than $2$ contiguous 
fragments. If a read is only a single contiguous string of alleles, the spatial location of that read is placed on the diagonal in $[n]^2$, i.e., the spatial label
assigned to the read is a pair of identical numbers (each being the starting position of the single read). If a read happens to consist of more than $2$ contiguous 
fragments, there are several possibilities when it comes to assigning it a spatial label. For instance, we may split the read into multiple reads such that each 
one has at most two contiguous fragments; alternatively, we may choose two of the multiple starting contiguous points in a suitable fashion to form the 
spatial label. Further details regarding assignment of spatial labels are discussed in Section \ref{sec:real_data}.

\subsection*{Summary of the Algorithmic Pipeline}

Our algorithm takes the spatial graph $G$ as input, and produces the reconstructed haplotypes. After pre-processing the sequencing data, we may form the graph $G$ and assign spatial labels $(X_u)_{u \in [n]}$ to its nodes. However, the community 
labels $(Z_u)_{u \in [n]}$ are unknown at this point. We defer a detailed discussion of the computational complexity involved in the pre-processing needed to 
construct $G$ to the end of this section, where we show that one can exploit the structure in the data to reduce complexity of this pre-processing step from 
the naive $O(n^2)$ to roughly $O(n)$. Given the representation of the data by the graph $G$, and having assigned spatial labeling to its nodes, our algorithm has two main components - {\em (i)} - a 
community detection on $G$ to discover origin of each read and thus effectively group the reads into $k$ different clusters and {\em (ii)} - Subsequently, for all $i \in [m]$ 
and $l \in [k]$, we estimate $\hat{s}_l[i]$ by a simple majority rule as described in the sequel.
\\

%to be the majority alphabet indicated by the reads covering site $i$ and are estimated to originate from string $l$ by  the community detection algorithm. 

\noindent {\bf 1. Euclidean Community Detection} - This part of the algorithm is where we take as input the graph $G$ along with the spatial labels $(X_u)_{u \in [n]}$ and estimate for each $u \in [n]$, a community label $\widehat{Z}_u \in [k]$, denoting which of the $k$ haplotype, a read is likely originating from.  We summarize the key steps in this task. The formal pseudo code is given in Algorithm \ref{alg:main_routine}.

\begin{enumerate}
	
	\item We first tessellate the grid $[n]^2$ into smaller \emph{overlapping} boxes, denoted by $(B_{x,y})_{1 \leq x \leq \tilde{n}, 1 \leq y \leq \tilde{n}}$. 
	Here $\tilde{n} < n$ is a parameter which we choose and each $B_{x,y} \subset [n]^2$. The tessellation is such that each grid point $u \in [n]^2$ belongs
	to multiple boxes $B_{x,y}$ since the boxes overlap.
	
	\item For each box $B_{x,y}$, let $H_{x,y}$ denote the subgraph of $G$ containing nodes whose spatial locations lie in $B_{x,y}$. The nodes of $H_{x,y}$ 
	are all clustered independently into $k$ communities.
	
	\item The community estimates in different boxes are \emph{synchronized} to obtain a global clustering estimate from spatially-local clustering estimates. 
	Since each grid point is present in multiple boxes, a read gets many estimates for its community, each of which adds an `evidence' to the label of the node. 
	This scheme has a natural `error-correcting' mechanism, since it is less likely for a node to be misclassified in the majority of the boxes it lies in, as opposed 
	to any one particular box.
	
%	\item Finally, 
	
\end{enumerate}

\noindent {\bf 2. Reduce by Majority} - After estimating for each node (read), the likely haplotype from which it originates $(\widehat{Z}_u)_{u \in [n]}$, we reconstruct the haplotypes by a simple majority vote. For all $j \in [k]$ and $i \in [m]$, we estimate $\hat{s}_j[i]$ to be the majority among the $4$ letters in the alphabet as indicated by the reads 
that cover site $i$ and are estimated to belong to string $j$ in the above clustering step.

\subsection*{Intuition behind the Algorithm}

\begin{figure*}[!tpb]
	\centering
	%\minipage{0.25\textwidth}
	\subfigure{
		\includegraphics[width=0.2\textwidth]{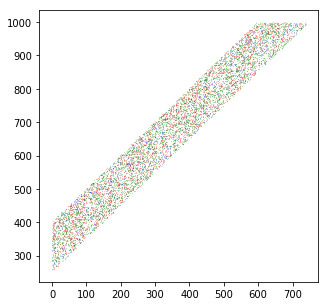}}
	%\endminipage\hfill
	%\minipage{0.25\textwidth}
	\subfigure{
		\includegraphics[width=0.2\textwidth]{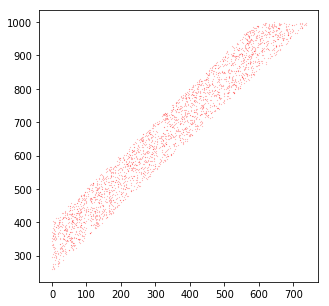}}
	% \endminipage\hfill
	%	\minipage{0.25\textwidth}
	\subfigure{
		\includegraphics[width=0.2\textwidth]{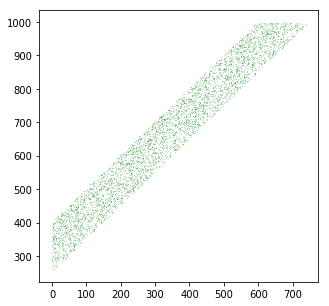}}
	% \endminipage\hfill
	% 	\minipage{0.25\textwidth}
	\subfigure{
		\includegraphics[width=0.2\textwidth]{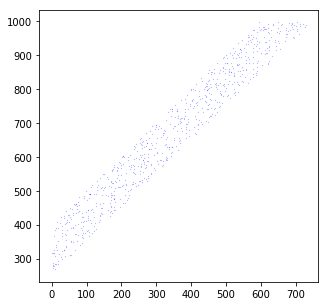}}
	% \endminipage\hfill
	
	\caption{An instance with three strings (haplotypes) of length $1000$ where spectral clustering on $G$ fails. The coverage is $10$, the effective read 
		lengths $r=2$ and $R=250$. The error probability $p=0.05$. The figure on the left is the union of the three figures on the right. The three colored plots 
		are the spatial locations of the recovered communities by the spectral algorithm applied on $G$. The density of the recovered blue estimates is $0.0995$, 
		while that of red is $0.33$ and of green is $0.57$. The total overlap achieved by the spectral method is $0.1$. Note that in the ground truth all three 
		colors are equal in intensity, which is not captured by the spectral method. However, our correctly algorithm predicts approximately equal-sized 
		communities, achieves an overlap of $0.98$ and runs 4 times faster.}
	\label{fig:spectral_bad}
\end{figure*}

Observe that for any two reads $u$ and $v$ that overlap,  if the weight $w_{uv}$ is positive and large, then they are likely from the same haplotype while if negative and large, they are likely from different haplotypes. Consider a subset of reads such that the 
absolute value of the weights on the edges that connect them in graph $G$ are `large'. This subset of reads can readily be grouped into up to $k$ different clusters
using standard techniques such as the spectral method. Such simple schemes will be successful in this scenario since the `signal' for clustering in the subgraph is high 
due to strong \emph{interaction} between the corresponding reads, i.e., the adjacency matrix of the subgraph is dense. 
\\

However, the entire set of reads does not 
posses the aforementioned property -- in particular, it has a large diameter (of order $n$). Thus, standard spectral methods applied to graph $G$ are both 
computationally impractical as well as statistically sub-optimal for recovery. The computational impracticality is easy to observe; indeed, any clustering scheme will be 
super-linear in the number of nodes $N_n$, which renders them extremely slow on large problem instances that are typical in practice. Furthermore, even the
pre-processing of reads to construct graph $G$ is of order $N_n^2$, which makes it computationally challenging in practical settings. The statistical sub-optimality is 
more subtle and stems from the fluctuations in the density of reads in space; in many problem instances, the density of reads varies across space due to randomness 
in the read generation process. For instance, in Figure \ref{fig:spectral_bad}, we see an example where the density of reads captured by the spectral algorithm is highly 
imbalanced due to the fluctuations of the nodes in space. However, in the ground truth set, the statistical distribution of reads across space is the same for all communities.
Therefore, to improve performance, one would need an additional `constraint' in the reconstruction algorithm to enforce the fact that the spatial distribution of reads 
across all communities is the same. 
\\

Our strategy in this paper is to first partition the set of reads into subsets wherein the reads interact strongly. Since the reads have 
spatial locality, we partition the set of reads into boxes as illustrated in Figure \ref{fig:boxes}. In each box, we consider the subgraph of $G$ with nodes having spatial 
labels lying in that box and then cluster this sub-graph independently of other subgraphs in other boxes. The partition of space into boxes is fixed a priori and is data 
independent. The box size and how much they overlap are hyper-parameters that can be tuned. In each box, we run a clustering algorithm and then combine the 
estimates from different boxes to form a final global partitioning of the nodes of $G$. The partitioning based on spatial locality automatically ensures that the spatial 
density of the estimated communities are roughly identical. The intuition for this stems from the fact that the reads will be roughly uniformly distributed within a 
box since the box is `small' in size. More importantly, by requiring that the boxes overlap, a single read will be present in multiple boxes. This further boosts statistical 
accuracy of clustering by embedding natural \emph{error-correction} scheme; since a single read is processed multiple times, there are multiple instances of `evidence' 
for the community label of a node. From a computational complexity viewpoint, partitioning the set of nodes and clustering smaller instances dramatically reduces 
run-time as a majority of clustering algorithms are super-linear in the number of data points and hence reducing the sizes of the graphs to be clustered has a significant 
beneficial impact. Therefore, our algorithm is both computationally feasible on large instances and is statistically superior compared to standard graph clustering 
algorithms directly applied on $G$.

 \subsection{Pseudo Code}
 
 Let us start by introducing the notation needed to formalize the algorithm. The algorithm has hyper-parameters $A,B, \text{\it iter},M \in \mathbb{N}$ and 
 $\alpha \in [0,1]$. For $x,y \in [ \lceil \frac{\tilde{n}}{A} \rceil ]$, we define $B_{x,y} \subset [n]^2$ as $B_{x,y} := [Ax, \min(Ax+B,\tilde{n})] \times [Ay, 
 \min(Ay+B,\tilde{n})]$, the box indexed by $(x,y)$. Thus, the parameters $A$ and $B$ dictate how large a  box is and how many boxes cover a read. In the 
 course of the algorithm, we maintain a dictionary of lists $\mathcal{C}$, where for each node $u \in [N_n]$, $\mathcal{C}[u]$ denotes the list of community 
 estimates for node $u$. Each node has more than one estimate as it belongs to more than one box. The estimates from clustering in each box are added as 
 `evidence' of the community estimate for the nodes in the box. Having multiple estimates for a node helps in combating clustering errors in certain boxes.
 \\
 
\begin{figure}[!tpb]
 \centering
 \includegraphics[scale=0.1]{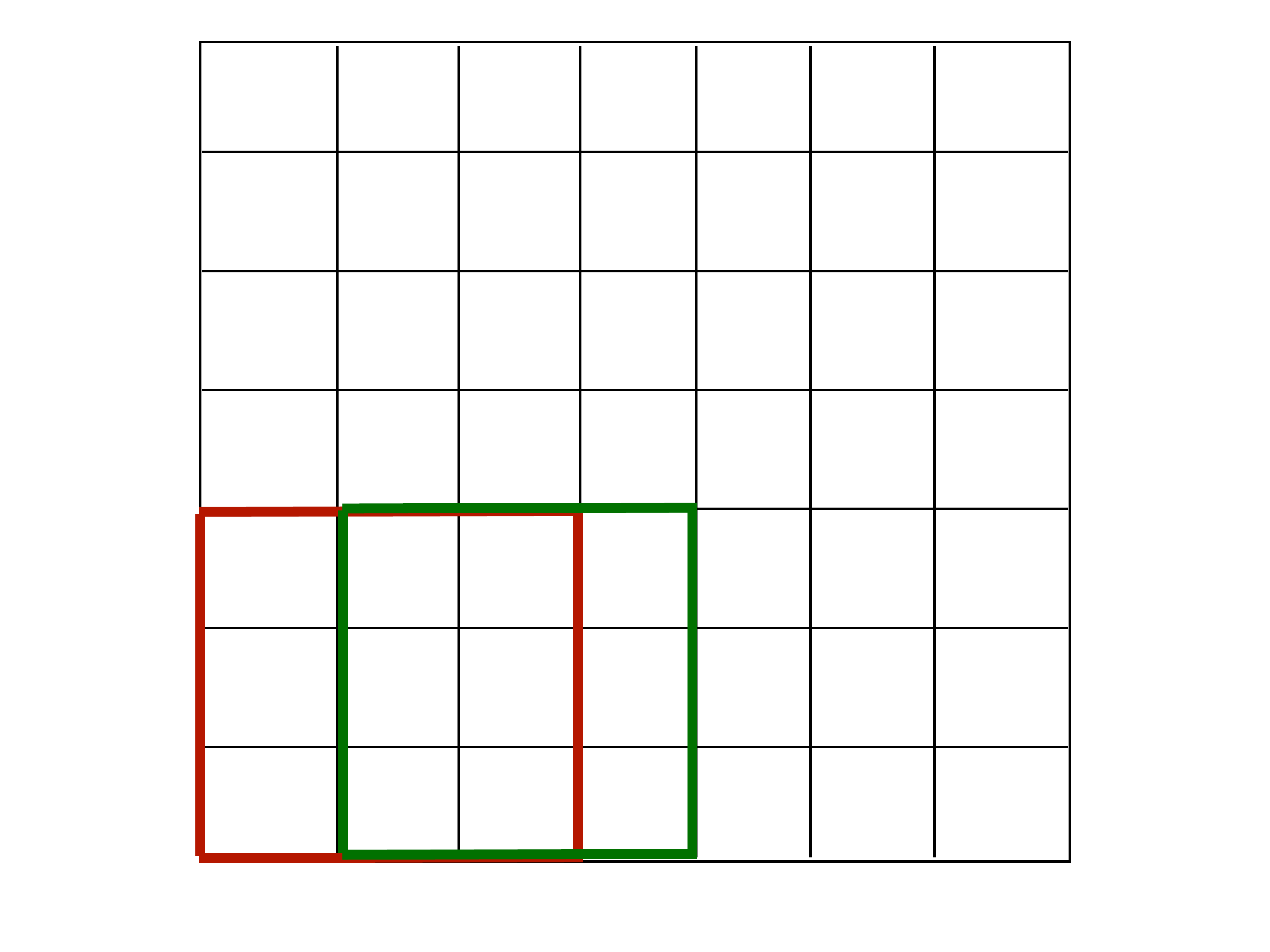}
 \caption{This is an example with $\tilde{n}=7$. The parameters $A=1$ and $B=3$. The red and green boxes are examples of $B_{x,y}$.}
 \label{fig:boxes}
 \end{figure}

 \noindent {\bf Main Routine} - We now describe the algorithm in detail. The first step consists of partitioning the space $[n]^2$ into multiple overlapping boxes as shown in 
  Figure \ref{fig:boxes}. Recall that the hyper-parameters $A$ and $B$ allow one to tune both the size of a box and the number of boxes that will cover a given 
  location of $[n]^2$. In each box indexed by $(x,y)$ for $x,y \in [ \lceil \frac{\tilde{n}}{A} \rceil ]$, we identify the nodes of $G$ having their spatial label in that 
  box; let us denote the collection of such nodes by $H_{x,y}$. If the number of nodes in $H_{x,y}$ is small (e.g., smaller than a certain hyper-parameter $M$),
  then we do not attempt to cluster these nodes. We need to set such a minimum size of $H_{x,y}$ or the output of clustering may turn out to be noisy and 
  non-informative. In addition, if more than an $\alpha < 1$ fraction of nodes in $H_{x,y}$ have at least one community estimate, then again we do not cluster 
  $H_{x,y}$. The reason for doing so is to decrease the running time by ensuring we only perform the clustering step when there are sufficiently many new unexplored 
  nodes. In each remaining box $(x,y)$ (i.e., each box with at least $M$ nodes where at most an $\alpha$ fraction of them have prior estimates) we apply a fast and 
  simple local clustering algorithm. In particular, we generate an approximate clustering of the nodes in $H_{x,y}$ by directly running a standard $k$-means 
  algorithm \cite{kmeans_alg} on the adjacency matrix of $H_{x,y}$. We then iteratively improve upon this initial guess by reassigning each node to the most likely 
  cluster while keeping the assignment of other nodes fixed. This iterative update rule is reminiscent of the class of Expectation Maximization algorithms, although 
  our method is fast and \emph{non-parametric}. We run the iterative clean-up procedure for {\ttfamily iter} number of iterations. Once the nodes of $H_{x,y}$ are 
  clustered, we append the result to the dictionary of lists $\mathcal{C}$ after appropriately synchronizing the community estimates of each node. Once we have 
  iterated over all boxes, we assign a single community estimate to each node based on the majority in the list of estimates in $\mathcal{C}$. The algorithm is
  formalized as the following pseudo-code.
  \\
 
 \begin{algorithm}[htb]
 	\normalsize
 	\caption{Main Routine}
 	\label{alg:main_routine}
 	\begin{algorithmic}[1]
 		\Procedure{Main}{$G,k,(X_{i})_{i \in [1,N_n]},A,B,\text{iter},M$} \Comment{$A, B, \text{iter}, M, \alpha$ are hyper-parameters}
 	%	\State Tessellate $[n]^2$. For $x,y \in \left[ \lceil \frac{n}{A} \rceil \right]$, let $B_{x,y} = [Ax,\min(Ax+B,n)] \times [Ay,\min(Ay+B,n)]$.
 	\For {$i \in \{1,\cdots,n\}$} \Comment{For every node(read)}
 	\State $\mathcal{C}[i] \gets []$ \Comment{Initialize to empty}
 	\EndFor
 		\For {$1 \leq x \leq \left[ \lceil \frac{n}{A} \rceil \right]$ }
 		\For {$1 \leq y \leq \left[ \lceil \frac{n}{A} \rceil \right]$ }
 		\State $B_{x,y} \gets [Ax,\min(Ax+B,n)] \times [Ay,\min(Ay+B,n)]$
 		\State $H_{x,y} \gets$ Subgraph of $G$ with spatial labels in $B_{x,y}$
 		\State $f \gets\frac{ |\{l \in H_{x,y} : |\mathcal{C}[l]| > 0|}{|H_{x,y}|}$ \Comment{Fraction of nodes with a community estimate} 
 		\If {$|H_{x,y}| \geq M$ {\bf {\ttfamily AND}}  $f \geq 1-\alpha$}
 		\State $\mathbf{e}$ $\gets$ {\ttfamily LOCAL-CLUSTER}($H_{x,y},k$, iter)
 		\State $\mathcal{C} \gets ${\ttfamily SYNCHRONIZE}($\mathcal{C},\mathbf{e}$)
 		\EndIf
 		\EndFor
 		\EndFor \\
 		\Return {\ttfamily REDUCE-BY-MAJORITY}($\mathcal{C}$) 
 	%	\State $(\widehat{s}_l[i])_{l \in \{1,\cdots,k\},i \in \{1,\cdots,m\}} \gets \text{{\ttfamily ESTIMATE-MAJORITY}}(\mathcal{C})$.
%\Return $(\widehat{s}_l[i])_{l \in \{1,\cdots,k\},i \in \{1,\cdots,m\}}$
 		\EndProcedure
 	\end{algorithmic}
 \end{algorithm}
 
 \begin{algorithm}[htb]
 	\normalsize
 	\caption{Small Graph Clustering}
 	\label{alg:local_clustering}
 	\begin{algorithmic}[1]
 		\Procedure{Local-Cluster}{$H,k$, iter}
 		\State $\tilde{\mathbf{e}} := \text{k-means-cluster}(H,k)$ \Comment{ $k$-means Algorithm of \cite{kmeans_alg}}
 		\State $\mathbf{e}_0 = \text{One-Hot-Encoding} (\tilde{\mathbf{e}})$.
 		\For {$1 \leq i \leq \text{iter}$}
 		\State $\mathbf{e}_i = \arg\max H\mathbf{e}_{i-1}$ \Comment {row wise argmax}
 		\EndFor \\
 		\Return $\mathbf{e}_{\text{iter}}$.
 		\EndProcedure
 	\end{algorithmic}
 \end{algorithm}

\noindent {\bf Local Clustering Step} - This step is described in Algorithm \ref{alg:local_clustering}. We follow a two step-procedure. In the first step, we get an approximate clustering of 
the graph $H$ by applying the standard $k$-means algorithm on the adjacency matrix $H$. We then one-hot encode this clustering result. One hot encoding is one where for each read we associate a $k \times 1$ vector in which all entries re $0$ except for a single entry corresponding to the estimated community label of that read to be $1$. More 
formally, if $r$ denotes the number of nodes of $H$, then the one-hot encoding result is a matrix $\mathbf{e}_0$ of size $r \times k$. Each entry 
of $\mathbf{e}_0$ is either $0$ or $1$; the entry in the $i$th row and $j$th column being $1$ implies that the $i$th node is classified as belonging 
to community $j$. Thus, each row of $\mathbf{e}_0$ contains exactly one $1$ while the rest of the entries are all $0$. We then run a `clean-up' 
procedure by iteratively updating the estimate as follows.
\begin{align}
\mathbf{e}_{t+1} = \mathcal{T}( H \mathbf{e}_t ).
\label{eqn:local_update}
\end{align}
The function $\mathcal{T}$ is applied row-wise; for matrix $A$, it sets the $i$th row and $j$th column of $\mathcal{T}(A)$ to $1$ if $j = \arg \max A[i]$, 
else the $i$th row and $j$th column of $\mathcal{T}(A)$ is set to $0$. If a row has more than one column where the maximum is attained, the first column 
where the maximum occurs is assigned value $1$ while the other columns are assigned value $0$. Hence the dimensions of $A$ and $\mathcal{T}(A)$ are 
the same. Furthermore, for any matrix $A$, the matrix $\mathcal{T}(A)$ is such that its entries are either $0$ or $1$, with each row having exactly one 
entry valued $1$. 
\\

The iterative update is based on the following intuition. Let the clustering be encoded by matrix $\mathbf{e}_t$ for some $t \in \mathbb{N}$, and consider 
a tagged node $u \in [n]$. The new updated value of the community label of node $u$ is then the `most-likely' label \emph{given} the estimates of the
community labels of the rest of the nodes. More precisely, the `weight' that a node $u$ is in a community $l \in [k]$ is the sum of the weights along the edges 
connecting $u$ to $v \in [n]$ in $G$ such that the estimate of node $v$ is $l$. The new community label of node $u$ is the one having the largest weight. 
By performing this operation simultaneously for all nodes, one obtains the representation in Equation (\ref{eqn:local_update}). The intuition for reassigning 
the node to the community with the maximum weight stems from the observation that if a weight along an edge is positive and large, then it is likely that the 
end nodes of the edge share the origin, i.e., the corresponding reads sample the same haplotype. Likewise, if the weight along an edge is negative and large, 
then it is likely that the end nodes represent reads that sample different haplotypes. Therefore, for the iterative update to perform well, the initial estimate 
$\mathbf{e}_0$ must be `good enough'; we achieve this by applying the k-means clustering algorithm on the adjacency matrix $H$. In principle, one can obtain 
somewhat better initial guess for $\mathbf{e}_0$ by applying the k-means algorithm to the eigenvectors of $H$, but the marginal gains in statistical accuracy 
does not warrant the enormous increase in computation needed to perform such a spectral clustering.
\\

The clean-up method, at first glance, seems to bear similarities to other dynamical algorithms such as expectation maximization, Belief Propagation (BP) and 
tensor factorization based methods of \cite{vikalo_tensor}. Unlike BP, however, we do not iterate the beliefs or probabilities of a node belonging to various 
communities; instead, we make a hard assignment at each update step. While for standard BP algorithms it is desirable that a graph is tree-structured, our graph
contains a lot of triangles and loops due to the spatial embeddings. Therefore, it would be insufficient to keep track of the node marginals -- instead, BP would
need the entire joint distribution which is not tractable. Despite the undesirable properties of $G$, benchmarking results demonstrate that our algorithm 
performs community detection on the graph very well.
\\

 \begin{algorithm}[htb]
 	\normalsize
	\caption{Synchronization Step}
	\label{alg:synchronize}
	\begin{algorithmic}[1]
		\Procedure{Synchronize}{$\mathcal{C}, \mathbf{e}$} 
		\State  $\widehat{W} \gets 0$
		\State  $\widehat{\pi}\gets \text{id}$ \Comment{The identity permutation}
		\For {All permutations $\pi$ of $[k]$}
		\State local-weight $\gets 0$
		\For {All nodes in $e$}
	%	\State local-weight $\gets$ local-weight $+ \text{fraction}( \mathcal{C}[\text{node}], \pi(\mathbf{e}[\text{node}])$.
		\State $W_{\pi} \gets \sum_{i \in \mathcal{N}_{x,y}} \mathbf{1}_{\mathcal{C}[i] \neq \text{empty}} \frac{\sum_{j \in \mathcal{C}[i]} 
			\mathbf{1}_{ \pi(\sigma[i]) =  \mathcal{C}[i][j]}   }{\sum_{j \in \mathcal{C}[i]}1}$
		\EndFor
		\If {$W_{\pi} >\widehat{W}$}
		\State $\widehat{W}\gets W_{\pi}$
		\State $\widehat{\pi}\gets \pi$.
		\EndIf
		\EndFor
		\For {nodes in $\mathbf{e}$}
		\State Append $\widehat{\pi}(\mathbf{e}[\text{nodes}])$ to $\mathcal{C}[\text{nodes}]$.
		\EndFor \\
		\Return $\mathcal{C}$.
		\EndProcedure
	\end{algorithmic}
\end{algorithm}

\begin{algorithm}[htb]
	\normalsize
	\caption{Reduce by Majority}
	\label{alg:reduce_majority}
	\begin{algorithmic}[1]
		\Procedure{Reduce-Majority}{$\mathcal{C}$} 
		\For {node in $\mathcal{C}$.keys}
		\State $\mathcal{C}[\text{node}] \gets \text{Majority}(\mathcal{C}[\text{node}])$. \Comment {Estimate haplotype by majority}
		\EndFor 
		
		\For {$l \in \{1,\cdots,k\}$}
		\For {$i \in \{1.\cdots,m\}$}
		\State $\widehat{s}_l[i] \gets$ majority of alphabets among reads $j$ with $\mathcal{C}[j] =l$ and covering site $i$.
		\EndFor
		\EndFor \\
		\Return $(\widehat{s}_l[i])_{l \in \{1,\cdots,k\},i \in \{1,\cdots,m\}}$
		\EndProcedure
	\end{algorithmic}
\end{algorithm}

\noindent {\bf Synchronization Step} - The main routine in Algorithm \ref{alg:main_routine} considers the boxes sequentially and performs local clustering steps. Once the 
local clustering is performed, a key component is to \emph{synchronize} the estimates of the current box with the estimates of the 
boxes that are already clustered. The synchronization is essential since the problem is a permutation invariant to the labels. Formally, 
the statistical distribution of the data remains unchanged if the true underlying labels of the $[k]$ strings are permuted. Hence, the 
best hope for any recovery algorithm is to reconstruct the $k$ strings upto a permutation of labels. Thus, if any clustering algorithm is 
run on two different subsets of nodes, the corresponding haplotype estimates need to be synchronized to produce a clustering of the 
nodes in the union of the sets. We perform this clustering in line $8$ of the main routine Algorithm~\ref{alg:main_routine} by invoking 
the sub-routine \ref{alg:synchronize}.
\\

In sub-routine  \ref{alg:synchronize}, we decide on how to permute the community label output of the local clustering estimate of 
$H_{x,y}$ that best `synchronizes' with the label estimates of the other nodes of $G$ at that instance. Observe that at the instant of 
synchronizing the output of $H_{x,y}$, other nodes of $G$ have either none or multiple label estimates. There is a possibility that more 
than one label estimate per node is present in multiple boxes, each adding an `evidence' for a node's cluster. We select a permutation of 
the labels by sequentially going over all permutations of $[k]$ and selecting the one that has the highest `synchronization-weight'. More 
formally, let $\mathcal{N}_{x,y} \subset [n]$ denote the indices of the nodes in $H_{x,y}$; for all $u \in \mathcal{N}_{x,y}$, denote by 
$\sigma[u] \in [k]$ the label estimates formed by the local clustering on $H_{x,y}$. The synchronization weight for a permutation $\pi$ of 
$[k]$ is defined as
\begin{align*}
W_{\pi}:=\sum_{i \in \mathcal{N}_{x,y}} \mathbf{1}_{\mathcal{C}[i] \neq \text{empty}} \frac{\sum_{j \in \mathcal{C}[i]} 
\mathbf{1}_{ \pi(\sigma[i]) =  \mathcal{C}[i][j]}   }{\sum_{j \in \mathcal{C}[i]}1}.
\end{align*}
In words, we go over all nodes in $H_{x,y}$ that have at least one prior estimate and sum the fraction of the previous estimate equaling 
the label assigned by the local clustering $H_{x,y}$ after applying the permutation $\pi$ to the local clustering's output. Among all permutations, 
we select the $\pi^{*}$ having the highest synchronization weight (ties are broken arbitrarily). After doing so, for each node $u$ of $H_{x,y}$
we append the label $\pi^{*}(\sigma(u))$ to the list $\mathcal{C}[u]$. The key feature of the above procedure is consideration of the 
\emph{fraction}, which is a proxy for the `belief' of the community label of a node, rather than just a count; this is meaningful because the counts 
across different nodes can be significantly skewed by the order in which the boxes are clustered and synchronized.

\subsection*{Computational Complexity}

In this section, we discuss the computational complexity of implementing our algorithm and the effect of various hyper-parameters on runtimes. 
A naive implementation of the algorithm would incur a cost of the order $n^2$ just to construct the graph $G$ from the reads. This step itself may 
be infeasible in practical scenarios where the number of reads will be on the order of millions. However, our algorithm only needs the subgraphs 
$H_{x,y}$ instead of the full graph $G$. Therefore, we pre-process the reads and create a hash-map where for each location in $[n]^2$, we store 
the list of reads that have spatial label in that location. This requires one pass through the list of reads, i.e., has computational complexity of order 
$n$ and storage complexity of order $n$. Now, creating the adjacency matrix $H_{x,y}$ is quadratic in only the number of nodes in $H_{x,y}$. The 
synchronization step requires time complexity of the order of the number of nodes in $H_{x,y}$ times the number of distinct permutations of $[k]$.

\subsection*{Choosing the Parameters of our Algorithm}

Our algorithm has an inherent trade-off between computational complexity and statistical accuracy that can be achieved by varying 
the hyper-parameters. For instance, if we decrease $A$ while keeping $B$ fixed, the number of boxes and therefore the computational time are
increased. However, the statistical accuracy would improve since each node would now be present in many boxes and hence the error-correction 
scheme performs more accurately. Similarly, increasing the parameter $M$ can reduce the run-time by considering fewer boxes to perform local 
clustering, while potentially decreasing statistical accuracy because there is less evidence for determining community label of each node.

\begin{table}[h!]
	\centering 
	\normalsize
	\begin{tabular}{cc |cccc |  cccc |cccc}
		\hline
		
		\multirow{2}{*}{Cov} & \multirow{2}{*}{Err} & \multicolumn{4}{c}{ComHapDet} &   \multicolumn{4}{c}{AltHap} &  \multicolumn{4}{c}{HPoP} \\ \cline{3-14}
		&&$\text{CPR}$ & MEC & t(s)& $\sigma$ & $\text{CPR}$ & MEC & t(s)& $\sigma$ & $\text{CPR}$ & MEC & t(s)& $\sigma$  \\\hline
		
		%& $0.3$ & $33.95$ & $2524.61$ & $5.03$ & $2.4$ & $52.19$      \\

		\multirow{3}{*}{$7$} & $0.05$ & $99.2$ & $\bf{662.7}$ & $18.3$ & $0.2$    & $\bf{99.9}$ & $960.7$ & $13.4$ & $0$ &$99.8$ & $961.5$ & $3.1$ & $0.1$  \\
		& $0.1$ & $98.2$ & $\bf{1289.1}$ & $18.8$ & $0.4$   & $\bf{99.8}$ & $1871.2$ & $13.8$ & $0.1$ &  $99.4$ & $ 1868.5$ & $ 3.4$ & $ 0.3$  \\
		& $0.2$ &$80.5$ & $\bf{2640}$ & $18.2$ & $1.6$ & $\bf{85.9}$ & $4844.1$ & $13.7$ & $1.3$ & $84.8$&$3862.7$ & $3.5$ & $8.6$     \\\hline

		\multirow{3}{*}{$10$ } & $0.05$ & $\bf{99.9}$ & $\bf{923.4}$ & $29.2$ & $0.1$  & ${99.9}$ & $1352.9$ & $15.43$ & $0$ &  ${99.9}$ & $1354.9$ & $1.7$ & $0$     \\
		
		& $0.1$ & $99.5$ & $\bf{1831.1}$ & $27.1$ & $5.3$  & $98.1$ & $3132.3$ & $15.5$ & $0.8$ &    $\bf{99.8}$ & $2667.5$ & $3.1$ & $0.4$     \\ 
		& $0.2$ & $91.9$ & $\bf{3575.9}$ & $27.9$ & $1.3$  & $\bf{92.8}$ & $5231.9$ & $24.2$ & $1.3$ &  $88.3$ & $5488.2$ & $3.3$ & $11.5$  \\
		%& $0.3$ & $42.58$ & $5327.6$ & $38.32$ & $4.93$ & $56.36$     \\
		\hline  
		\multirow{3}{*}{$15$ } & $0.05$ & $\bf{100}$ & $\bf{1382.7}$ & $52.1$ & $0$  & $100$ & $2034.5$ & $30$ & $0.1$ & ${100}$ &$2022.5$ & $8$ & $0$   \\
		& $0.1$ & $\bf{99.9}$ & $\bf{2772.9}$ & $56.4$ & $0.1$  & $99.9$ & $3989.7$ & $39.1$ & $0$ &$99.9$ & $3986.5$ & $7.3$ & $0$\\
		& $0.2$ & $\bf{97.9}$ & $\bf{5283.6}$ & $50$ & $0.4$  & $96.8$ & $7646.3$ & $39.2$ & $0.6$ & $96.7$ & $7789$ & $7$ & $1.8$ \\
		\hline

	\end{tabular} 
	
	\caption{Simulated diploid biallelic data.}
	\label{tab:diploid_read_len2_comdet_althap}
\end{table}

%{p{0.75cm}p{0.75cm}| p{0.7cm} p{0.7cm} p{0.7cm} p{0.7cm} p{0.7cm} |  p{0.7cm} p{0.7cm} p{0.7cm} p{0.7cm} p{0.7cm}}        	{ cc | ccccc |  ccccc }

\begin{table*}[h!]
	\centering
		\normalsize
			\begin{tabular}	{ cc | ccccc |  ccccc }
			\hline
			\multirow{2}{*}{Cov } & \multirow{2}{*}{Err} & \multicolumn{5}{c}{ComHapDet} &   \multicolumn{5}{c}{AltHap} \\ \cline{3-12}  
			&&$\text{CPR}$ & MEC & t(s)& $\sigma$ & M-CPR&$\text{CPR}$ & MEC & t(s)& $\sigma$ & M-CPR \\

		\hline

		\multirow{3}{*}{$7$ } & $0.002$ &$\boldsymbol{98.6}$ & $\bf{97}$ & $76.7$ & $0.9$ & $99.5$ & $89$ & $687$ & $295.2$ & $14$ & $93$      \\
		
		& $0.01$ &$\boldsymbol{93.8}$ &$\bf{662.1}$ & $81.2$ & $10.8$ & $97$ & $88.7$ & $966.2$ & $289.8$ & $17.5$ & $92.4$      \\
		& $0.05$ & $\boldsymbol{97.1}$ & $\bf{1504.7}$ &$75.5$ & $1.6$ & $98.9$ & $80.1$ & $2887.4$ & $332.1$ & $20.2$ & $86.3$      \\
		%\cline{3-12}
		\hline
		\multirow{3}{*}{$10$ } & $0.002$ & $\boldsymbol{99.8}$  & $\bf{93.7}$ & $137.5$ & $0.17$ & $99.9$ &$83.7$ & $1215.4$ & $593.2$ & $20.7$ & $88.4$      \\
		& $0.01$ & $\boldsymbol{99.7}$ & $\bf{413.1}$ & $135.9$ & $0.2$ & $99.9$ & $92.7$ & $1029.1$ & $592.7$ & $14.6$ & $95.4$     \\
		& $0.05$ & $\boldsymbol{99.4}$ & $\bf{2021.9}$ & $139.8$ & $0.3$ & $99.8$ & $92.7$ & ${3632.0}$ & $592.4$ & $14.6$ & $95.4$     \\
		%	\cline{3-12}
		\hline
		\multirow{3}{*}{$15$ } & $0.002$ & $\bf{99.9}$ & $\bf{124.6}$ &$300.4$ & $0.1$ & $99.9$ & $89.9$ & $1725$ & $708.5$ & $16.1$ & $94$      \\
		& $0.01$ & $\bf{99.9}$ & $\bf{611.1}$ & $307.9$ & $0.1$ & $99.9$ & $96$ & $1628.6$ & $ 781$ & $ 9.82$ & $ 97.6$   \\
		& $0.05$  & $\bf{99.9}$ & $\bf{2981.5}$ & $297.2$ & $0.2$ & $99.9$ & $87.4$ & $6721.3$ & $713.3$ & $20.4$ & $92.1$      \\
		\hline 
		
	\end{tabular}
	
	\caption{Simulated Triploid Tetraallelic Data. }
	\label{tab:triploid_ployallelic_simulated}
\end{table*}

%	{p{0.75cm}p{0.75cm}| p{0.7cm} p{0.7cm} p{0.7cm} p{0.7cm} p{0.7cm} |  p{0.7cm} p{0.7cm} p{0.7cm} p{0.7cm} p{0.7cm}}

\begin{table*}[htb!]
	\centering
		\normalsize
	\begin{tabular}{ cc | ccccc |  ccccc }

		\hline
		\multirow{2}{*}{Cov } & \multirow{2}{*}{Err} & \multicolumn{5}{c}{ComHapDet} &   \multicolumn{5}{c}{AltHap} \\ \cline{3-12}  
		&&$\text{CPR}$ & MEC & t(s)& $\sigma$ & M-CPR&$\text{CPR}$ & MEC & t(s)& $\sigma$ & M-CPR \\
		
		\hline

		\multirow{3}{*}{$7$ } & $0.002$ & $\bf{80}$ & $\bf{1316.3}$ & $143.5$ & $20.3$ & $91.8$ & $76.1$ & $1388.6$ & $521.4$ & $20.8$ & $87.5$     \\
		& $0.01$ & $79.1$ & $\bf{1640.0}$ & $118.5$ & $17.8$ & $91.8$ & $\bf{79.9}$ & $1812.8$ & $515.8$ & $20.5$ & $88.1$     \\
		& $0.05$ & $68.3$ & $3722.8$ & $129.7$ & $14$ & $87.3$ & $\bf{83.6}$ & $\bf{3481.9}$ & $503.1$ & $20.2$ & $92$      \\
		%\cline{2-13}
		\hline
		
		\multirow{3}{*}{$10$ } & $0.002$ & $\bf{98.9}$ & $\bf{193.1}$ & $253.3$ &$1.4$ &$99.6$ & $71.9$ & $1979.7$ & $594.3$ & $15.5$ & $85.6$     \\ 
		&$0.01$  & $\bf{99.1}$ & $\bf{585.9}$ & $261.8$ & $0.4$ & $99.8$ & $85.4$ & $1779.4$ & $585$ & $18.5$ & $92.1$     \\
		& $0.05$ & $\bf{98.2}$ & $\bf{2727.7}$ & $238.6$ & $0.6$ & $99.5$ &$78.6$ & $5331.4$ & $667.5$ & $15.6$ & $ 89.7$  \\
		%\cline{2-13}
		\hline
		\multirow{3}{*}{$15$} & $0.002$ &$\bf{99.8}$ & $\bf{182.7}$ & $487.0$ & $0.2$ & $99.9$ & $85.2$ & $ 2614.6$ & $ 684.5$ & $18.4$ & $92$ \\
		& $0.01$ & $\bf{99.8}$ &$\bf{806.5}$ & $482.7$ & $0.2$ & $99.9$ & $83.5$ & $3973.7$ & $684.1$ & $17.4$ & $92.6$ \\
		& $0.05$ & $\bf{99}$ & $\bf{4101.4}$ & $523.8$ & $298.9$ & $99.7$ &$95.1$ & $6397.6$ & $682.5$ & $14.5$ & $97.4$ \\
		\hline

	\end{tabular}
	
	\caption{Simulated Tetraploid Tetraallelic Data.   }
	\label{tab:tetraploid_ployallelic_simulated}
\end{table*}

%	{p{0.75cm}p{0.75cm}| p{0.7cm} p{0.7cm} p{0.7cm} p{0.7cm} p{0.7cm} |  p{0.7cm} p{0.7cm} p{0.7cm} p{0.7cm} p{0.7cm}}

\begin{table*}[htb!]
	\centering
	\normalsize
	\begin{tabular}{ cc | ccccc |  ccccc }

		\hline
		\multirow{2}{*}{Cov } & \multirow{2}{*}{Err} & \multicolumn{5}{c}{ComHapDet} &   \multicolumn{5}{c}{AltHap} \\ \cline{3-12}  
		&&$\text{CPR}$ & MEC & t(s)& $\sigma$ & M-CPR&$\text{CPR}$ & MEC & t(s)& $\sigma$ & M-CPR \\
		\hline
		
		%		\multirow{3}{*}{$5$ } & $0.002$ & $6$ & $7718.6$ & $165.07$ & $1.79$ & $52.49$ & $00$ & $000$ & $000$ & $000$ & $00$ \\
		%		& $0.01$  & $4.78$ & $8116.7$ & $154.91$ & $1.61$ & $51.23$ \\
		%		& $0.05$ & $2.53$ & $9541.4$ & $146.62$ & $0.51$ & $47.07$ \\
		%\cline{2-13}
		%		\hline
		\multirow{3}{*}{$10$ } & $0.002$ & $\bf{78.9}$ & $2256.6$ & $551.1$ & $15.6$ & $94.1$ & $76$ & $\bf{2022.9}$ & $977.9$ & $20$ & $90.6$     \\
		& $0.01$ &$\bf{84.1}$ & $\bf{2250.4}$ & $563.2$ & $14$ & $95.8$ & $70.4$ & $3533.7$ & $919.9$ & $19.9$ & $86.8$       \\
		& $0.05$ & $ 48.8$ & $9578.4$ & $526.3$ & $25.6$ & $81.9$  & $\bf{75.8}$ & $\bf{7440.7}$ & $1222.1$ & $17.9$ & $90$     \\
		%\cline{2-13} 
		\hline
		\multirow{3}{*}{$15$ } & $0.002$ & $\bf{99.3}$ & $\bf{308.2}$ & $1295.6$ & $0.3$ & $99.9$  & $70.4$ & $4960.6$ & $1780.4$ & $25.2 $ & $87.3$    \\
		& $0.01$ & $\bf{97.4}$ & $\bf{1528.5}$ & $1359.1$ & $5.4$ & $99.4$  & $77.7$ & $5493.4$ & $1624.6$ & $23.2$ & $89.9$        \\
		& $0.05$ & $\bf{94.7}$ & $\mathbf{6554.2}$ & $1207.5$  & $11.7$ & $98.7$ & $65.9$ & $13751.6$ & $2406.3$ & $19$ & $87.2$       \\
		%\cline{2-13} 
		\hline
		\multirow{3}{*}{$20$ } & $0.002$ & $\bf{99.5}$ & $\bf{382.8}$ & $2097.1$ & $0.2$ & $99.9$  & $77.1$ & $7095.1$ & $7561.2$ & $19.3$ & $91.9$        \\
		& $0.01$ & $\bf{99.5}$ & $\bf{1654.3}$ & $2116.5$ & $0.2$ & $99.9$ & $87.3$ & $5905.4$ & $6862.1$ & $18$ & $96.1$        \\
		& $0.05$ & $\mathbf{99.6}$ & $\bf{7912.8}$ & $2298.9$ & $0.2$ & $99.9$ & $65.1$ & $23381.8$ & $8563.4$ & $24.5$ & $86.9$          \\
		\hline

	\end{tabular}
	
	\caption{Simulated Hexaploid Tetraallelic Data.  }
	\label{tab:hexaploid_ployallelic_simulated}
\end{table*}

\section{Experimental Evaluation }
\label{sec:experiments}

We evaluate the performance of our proposed algorithm on both simulated and experimental data. We implemented our algorithm in Python. The simulations, 
as well as the experimental evaluations were conducted on a single core Intel I5 Processor with $2.3$Ghz processor and $8$ GB $2133$ MHz LPDDR3 RAM. 

\subsection{Performance on Simulated Data}

 We first test the performance of our algorithm in simulations for both the diploid biallelic case as well as the more challenging polyploid polyallelic case. Since
 the ground truth in simulations is known, we use CPR, MEC and M-CPR as the primary performance benchmarks. The CPR and M-CPR are reported as percentages,
 for ease of presentation. We compare the performance over a range of problem parameters, namely the ploidy and alphabet size, as well as the measurement 
 parameters, in particular the coverage, average read length and error rates. {\color{black}In each case, the hyper-parameters were set to $A = 15$ and $B = 4$. The parameter $\alpha = 0.95$, 
 for all polyploid cases and $\alpha = 0.85$ in the diploid case.} Recall that the parameter $\alpha$ allows one to control  the
 trade-off between the run time and statistical accuracy; specifically, a lower value of $\alpha$ results in faster run times at the cost of reduced CPR. The column 
 $\sigma$ displays the standard deviation of CPR (after being multiplied by $100$, for consistency). 
 {\color{black}In each table and metric, the boldfaced entry 
 represents the algorithm with the best performance for that entry.  }
 
 %As the ploidy increases, the problem gets harder; consequently, the run times increase andCPR deteriorates. We also notice that with increasing $r$,  the average read-length of a contiguous segment of a paired-end read, our algorithm performs much  better and achieves higher CPR while requiring shorter run times.

\subsubsection*{Simulated Data - Diploid Bialllelic Case}

In the diploid case, we rely on the synthetic paired-end read data used in \cite{vikalo_tensor}.  The average length of the effective (i.e., haplotype-informative) paired-end read $4$ with an insert gap in a paired end reads being uniformly sampled between $50$ to $150$. We use a haplotype length ($m$ in our notation) of $700$ in all case to be consistent with prior literature. As an example  \cite{diploid_simulated_benchmark}, which is often used to benchmark haplotype assembly methods. The 
results are reported in Table \ref{tab:diploid_read_len2_comdet_althap}. We simulate $15$ instances for each configuration of coverage and error probability and report the average in Table \ref{tab:diploid_read_len2_comdet_althap}. We compare our methods against AltHap 
\cite{vikalo_tensor}, a sparse tensor factorization method, and HPop \cite{h_pop}, a state-of-the-art dynamic programming approach to haplotype assembly. Since for diplod case CPR and M-CPR are identical, we only report the CPR.  
We 
restrict our attention to these methods since it is already established in \cite{vikalo_tensor}, \cite{h_pop}, \cite{superior}, that they are superior, both in terms of 
accuracy and run times, as compared to various other approaches including SDHaP \cite{sdhap}, an approach inspired by semi-definite programming relaxations 
of the max-cut problem, BP \cite{bp_hap}, a communications system design inspired belief propagation algorithm, HapTree \cite{hap_tree}, an algorithm inspired 
by a Bayesian reformulation of the problem, and HapCompass \cite{hap_compass}, an algorithm focused on finding cycle basis in a graphical representation of the 
haplotype assembly problem.

%However, as we do not generate vcf files in simulation, there is no direct way to apply the HapCompass method on a SNP fragment matrix and hence we do not report the performance of HapCompass in simulated data. 

%\subsubsection*{Simulated Data - Polyploid Biallelic Case}
%
%We next benchmark the polyploid haplotype assembly performance using the synthetic data in \cite{vikalo_tensor}. We compare the performance of our method with 
%AltHap, HPoP and HapCompas. We focus on these algorithms since it has been established in \cite{vikalo_tensor} that they outperform other popular polyploid biallelic 
%schemes such as SDhaP and BP \cite{bp_hap}. 
%

\subsubsection*{Simulated Data - Polyploid Polyallelic Case}

 We report the results in 
Tables \ref{tab:triploid_ployallelic_simulated}, \ref{tab:tetraploid_ployallelic_simulated}   and \ref{tab:hexaploid_ployallelic_simulated} for 
the cases of triploid, tetraploid and hexaploid, respectively. In all cases, we considered the tetra-allelic case, i.e., the case of alphabets of
size $4$, and the average length of the effective (i.e., haplotype-informative) paired-end read $4$. The average insert size between the 
paired end reads was chosen to be $200$ with a minimum gap of $50$. The benchmarking algorithm we consider is that of \cite{vikalo_tensor}, the state-of-the-art algorithm capable 
of polyploid polyallelic phasing; all other methods are restricted to biallelic variants. In each case, we test our algorithm on $10$ different
problem instances, where in each instance, a haplotype sequence of length $1000$ was phased. We use the same method and the publicly 
available code from \cite{vikalo_tensor} to generate the synthetic data for the various instances. For the hexaploid case, we do not report 
performances on coverage smaller than $10$, since the performance of both algorithms are poor.

\begin{table}
	\centering
	\normalsize
	\begin{tabular}{ |c|cc| }
	\hline Method & MEC Score  & t(secs) \\ \hline
	ComHapDet & $17738$ & $207$ \\
	AltHap & $14580$ & $105$ \\
	HPoP & $10596$ & $102$\\
	HapCompass & $12497$ &$375$ \\
	HapTree &$46617$ & {\color{black}$215$}\\
	\hline
	\end{tabular}
\caption{The performance comparison of the various algorithms on the biallelic tetraploid Potato dataset}
\label{tab:real_data}
\end{table}

\subsection{Performance on Real  Data}
\label{sec:real_data}

\subsubsection*{Tetraploid Potato Data Set}

We test our algorithm on a tetraploid real data set of Chromosome $5$ of Potato species \emph{Solanum Tuberosum}, whose reference genome 
is available publicly \footnote{ {\ttfamily http://ftp.ensemblgenomes.org/pub/plants/release-41/fasta/\\solanum\_tuberosum/dna/}}.
We considered a set of paired-end sequence reads reported in experiment {\ttfamily SRX3284127} available in the NCBI database 
\footnote{{\ttfamily https://www.ncbi.nlm.nih.gov/Traces/study/?acc=SRP119957}}. We then mapped the reads to the reference genome using 
the {\ttfamily BWA} software of \cite{bwa}. Subsequently, we use the {\ttfamily FreeBayes} software \cite{freebayes} for SNP calling and create the vcf file. We then extracted out connected components of reads and considered the haplotype assembly on instances that were at-least $20$ haplotypes long.
%After SNP calling, we identified a total of $4433$ haplotypes which were covered with a total of $25128$ reads. The SNPs are such that 
%they are split into $157$ different problem instances, where each instance has at least $20$ haplotypes long. 
%The SNPs in a single instance formed a connected component of reads, as described in Section \ref{sec:model}. 
%This leads to $157$ different problem instances (i.e., $157$ contiguous SNP subsequences) on which we benchmark the performance of various methods. 
The data set of reads after SNP calling is available in the  Github link provided in Abstract. We compare the performance of our algorithm with AltHap, 
HapCompass \cite{hapcompass}, HPoP \cite{h_pop} and HapTree \cite{hap_tree}, and report the results in Table \ref{tab:real_data}. We use the recommended hyper parameters of $A = 20, B=5, \alpha = 0.85$ and the minimum problem size as $20$. We chose these set of parameters to minimize run time, while at the same time ensuring that all reads in a data set are covered by at-least one community estimate.
\\

{\color{black} Table \ref{tab:real_data} compares MEC scores achieved by our method with those achieved by competing techniques. Note that 
the true accuracy is captured by the correct phasing rate but the ground truth data in this set is not available and thus CPR cannot be computed. 
While being a convenient surrogate metric, the MEC score may be misleading since e.g. a very low MEC score does not necessarily imply high 
CPR (see Table $1$ of \cite{vikalo_tensor} for an illustrative example). Note that the task of tetraploid phasing presents a challenge to our scheme
because the weight of an edge in the spatial graph (see Equation (\ref{eqn:weights_defn}) is biased towards being positive, even if the reads on 
the end points of an edge originate from different strings; this limits accuracy of assembly schemes that rely on read clustering. Finally, as 
illustrated by simulations, our methodology is suitable to settings where sequencing coverage exceeds $\sim10 \times$ (both in diploid and 
polyploid setting). At the time of writing this paper we do not have access to data sets with coverage beyond $10 \times$ and thus do not 
perform further experiments on real data.}

%As can be seen in the table, our algorithm performs favorably both in terms of  reconstruction accuracy expressed by the MEC score as well as in terms of the run time. However, we cannot evaluate CPR or M-CPR since the ground truth is not known. 

%
%\subsubsection*{1000 Genomes Project}
%
%We compare the performance of our algorithm on the 1000 Genomes project data from \cite{1000_genomes}. This is a dataset of human haplotypes and thus consists of diploid biallelic data. On this data-set, we also evaluate the achieved CPR by assuming a reference ground truth as given by \cite{ground_truth_diploid}. 

%\subsubsection*{Fosmid Dataset}
%
%The Fosmid data set comprises of longer set reads and higher ratio of SNPs to positions. In this data-set, we take the results of \cite{ground_truth_diploid} as reference ground truth against which we compare the CPR and M-CPR. 

\subsection{Results and Discussions}

The results indicate that our method is comparable to the state of the art, both in the diploid as well as the more challenging polyploid polyallelic scenario. In the diploid biallelic case, we see in Table  \ref{tab:diploid_read_len2_comdet_althap} that our method performs comparable to both AltHap and HPoP in terms of reconstruction accuracy as measured by CPR and MEC scores. In the polyploid polyallelic scenario however, tables \ref{tab:triploid_ployallelic_simulated}, \ref{tab:tetraploid_ployallelic_simulated} and \ref{tab:hexaploid_ployallelic_simulated} indicate that our algorithm is superior in terms of both CPR and MEC compared to the state of art, namely AltHap, which is the only methodology prior to our work that is capable of handling polyploid polyallelic data.  In terms of experiment on a real data ployploid experiment, we see in Table \ref{tab:real_data}, that our method performs comparably both in terms of reconstruction accuracy as measured by MEC score and runtime complexity as compared to other state of art methods. The results demonstrate that our methodology has significantly higher reconstruction at higher coverages as compared to lower coverages. This is unsurprising, as at higher coverages, there are more reads and hence more data to recover the haplotypes. Moreover, the runtime complexity of our method scales very gracefully with increasing coverage, making it attractive for many practical high coverage scenarios.

\section{Conclusions}

In this paper, we propose a novel methodology to assemble both diploid and ployploid haplotypes. The main observation we make is that, by a spatial representation of the paired-end reads, we can effectively convert the problem about haplotype assembly into a community detection task on a spatial graph. Our algorithm assigns to each paired end read, a spatial label corresponding to the starting indices of the two read fragments. We then divide the problem into overlapping instances, each of which considers the set of reads located nearby in this embedding and performs a community clustering, where the community label of a read (node) is the haplotype from which it originates from. Finally, for each read, we take the majority of the estimated communities from the various instances as the final community estimate of that read. We then use this estimated community labels for the reads to output the reconstructed haplotype.
\\

%\begin{backmatter}
	
%	\section*{Abbreviations} BP: Belief propagation, CPR: Correct phasing rate, M-CPR: Modified Correct phasing rate, MEC: Minimum error correction, SNP: Single nucleotide polymorphisms
	
%		\section*{Availability of Data and Materials}
%		All data are available on request. The code is available in \url{https://github.com/abishek90/ComHapDet-Repo}
%	
%
%	
%	\section*{Author's contributions}
%Algorithms and experiments were designed by AS, HV, and FB. Algorithm
%code was implemented and tested by AS. The manuscript was written
%by AS, HV and FB. All authors read and approved the final manuscript.

\textbf{Acknowledgements}

AS acknowledges numerous discussions with Abolfazel Hashemi and Ziqi Ke on this and related topics that helped in the conducting the numerical experiments. The authors also thank anonymous reviewers of ACM CNB MAC 2019, whose comments greatly helped shape the presentation. This work was funded in part by the NSF grant CCF 1618427 and a grant of the Simons foundation $\#$197982 to The University of Texas at Austin. 
	
%\section*{Funding}

%\section*{Acknowledgements}
%
%
%
%\section*{Ethics approval and consent to participate}
%Not applicable.
%
%\section*{Consent for Publication}
%Not applicable.
%
%	\section*{Competing interests}
%The authors declare that they have no competing interests.

\bibliographystyle{bmc-mathphys} 
%\bibliography{bmc_article}   

\bibliography{haplotype_BMC_Arxiv}

\end{document}